# BriFiSeg: a deep learning-based method for semantic and instance segmentation of nuclei in brightfield images

Mathieu Gendarme[1,2], Annika Lambert[1,3], Bachir El Debs[1,4]

[1] BioMed X Institute, Heidelberg, Germany

Code:  https://github.com/mgendarme/BriFiSeg
       https://github.com/mgendarme/nnUNet4BriFiSeg
Data:  https://doi.org/10.5281/zenodo.7195636

Authors contribution:
[2] Conceived, designed, and conducted the study, and wrote the manuscript
[3] Performed the wet lab experimental work
[4] Gave guidance and wrote the manuscript
correspondence to gendarme@bio.mx

Abstract: Generally, microscopy image analysis in biology relies on the segmentation of individual nuclei, using a dedicated stained image, to identify individual cells. However stained nuclei have drawbacks like the need for sample preparation, and specific equipment on the microscope but most importantly, and as it is in most cases, the nuclear stain is not relevant to the biological questions of interest but is solely used for the segmentation task. In this study, we used non-stained brightfield images for nuclei segmentation with the advantage that they can be acquired on any microscope from both live or fixed samples and do not necessitate specific sample preparation. Nuclei semantic segmentation from brightfield images was obtained, on four distinct cell lines with U-Net-based architectures. We tested systematically deep pre-trained encoders to identify the best performing in combination with the different neural network architectures used. Additionally, two distinct and effective strategies were employed for instance segmentation, followed by thorough instance evaluation. We obtained effective semantic and instance segmentation of nuclei in brightfield images from standard test sets as well as from very diverse biological contexts triggered upon treatment with various small molecule inhibitor. The code used in this study was made public to allow further use by the community.

## 1 INTRODUCTION

### 1.1 Classical image analysis in biology

To extract relevant information from microscopic cell images the first step consists in thresholding an image stained for nuclei [1]. The distinction between the foreground and background class, referred to as the semantic segmentation step, is applied and individual instances, using instance segmentation step, can then be derived using methods such as watershed [2]. Then measurements on single objects obtained can be performed to enable comparison between different conditions (Figure 1).

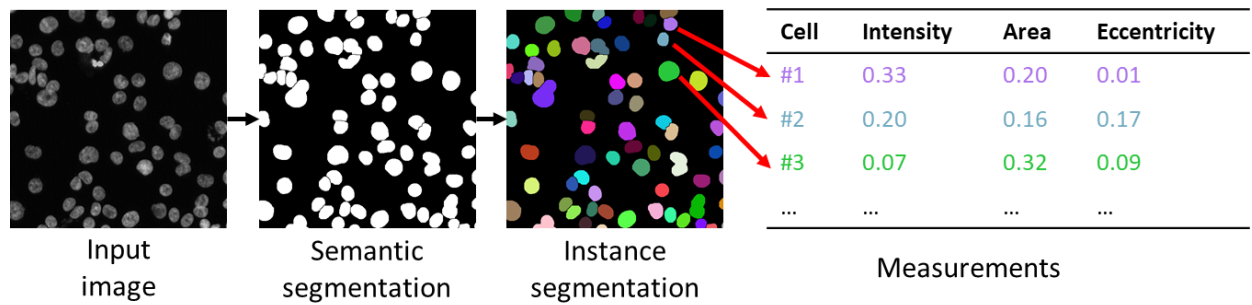

**Figure 1: Workflow for biological image analysis**
Pipeline describing the steps from the input image to the measurements performed on the individual nuclei. First the input image is thresholded to distinguish between the nuclei class and the background followed by the derivation of instances from the thresholded image and eventually measurements are performed on those individual objects.

1.2     Non-stained nuclei segmentation using deep learning-based method

The ability to segment individual nuclei without a dedicated stain opens new possibilities. First and foremost, images from live and fixed samples can be directly acquired without further sample preparation. Simpler and faster image acquisition can therefore be performed for applications that focus specifically on cell count and cell position such as cytotoxicity assay or cell migration assays respectively. Additionally, and for applications focusing on the analysis of fluorescence images, the nuclei dedicated channel can be employed for another and more biologically relevant marker. By doing so the staining capacity is increased by twenty five percent on standard 4-channels fluorescence microscope. For costly, time intensive or applications where material is limited, a stainless nuclei segmentation method would be highly beneficial to extract as much information as possible from the precious samples using high-content approaches. For all the reasons mentioned above we investigated the performance of deep learning-based methods to segment images from non-stained cell images.

Brightfield imaging was used to acquire images, because of its generic nature and availability on any microscope independently from the vessel format in comparison to phase contrast or differential interference contrast microscopy technique. The ability to perform complex semantic segmentation in the biomedical domain using deep learning based-methods has been already performed [3][4] in addition to the ability to segment nuclei from labelled images [5] or unlabelled images [6] was also shown. In this study we systematically addressed the impact of deep pre-trained encoders on semantic segmentation performance and benchmarked our method against others, showing that ours improved the segmentation performance. We also thoroughly investigated the instances generated from the semantic maps obtained prior.

In this study we focused on two distinct machine learning architectures, U-Net (Figure 3) [3] and FPN (Figure 4) [7], as they both proved to be relevant for nuclei semantic segmentation [4][8][9]. We systematically replaced the encoder of our architectures with classification convolutional neural network (CNN) trained on the ImageNet dataset and tested the effect on semantic segmentation performance. To our knowledge, transfer learning as described in [10][11] was not investigated in the context of semantic segmentation task in brightfield images. After optimizing the best architecture/encoder combination, we benchmarked it against the results published by Fishman *et al.* [6] and against the nnU-Net method [4] and showed that our approach improves performances in both cases.

We complemented our semantic segmentation method with two simple yet effective instance segmentation strategies using the semantic maps as input and finally investigated the quality of the instances generated as suggested by Caicedo *et al.* [8].

## 2 METHOD

### 2.1 Sample preparation and Image acquisition

RPE1, A549, HeLa and MCF7 cells were seeded (in 384-well plate, IBL Baustoff + Labor GmbH, #220.240.042) at 4000, 6000, 6000 and 8000 cells per well density respectively in 40 µL per well. Medium composition for RPE1 was DMEM/F-12, GlutaMAX supplement (Gibco, #31331028), 1x non-essential amino acids (Gibco, #11140035) and 10 % FBS (Gibco, #10270106). For all the other cell lines the medium composition was DMEM high glucose (Gibco, #10270106), non-essential amino acids (Gibco, #11140035) and 10 % FBS (Gibco, #10270106)). After seeding, the plates were placed and left for 24 h at 37◦C.

For staining, cells were fixed with 4 % PFA (Science Services, E15710) and incubated for 20 min. Plates were washed one time with PBS (5 min). Permeabilization of cells was done for 20 min (Triton X-100 (Amresco, 9002-93-1) diluted in PBS 0.1 % (V/V)). Plates were washed three time using PBS (5 min) and stained using Hoechst (1:10000 dilution in PBS, Life Technologies, #33342) and Phalloidin (1:1000 dilution in PBS, Thermo Fisher Scientific, #A12380), at RT for one hour. Plates were washed 3 more times using PBS (5 min). To prevent evaporation plates were sealing with transparent while allowing simultaneously brightfield imaging. Image acquisition was performed with a Nikon Ti2 microscope (CMOS camera DS-Qi2) using a 20x objective (CFI P-Apo DM 20x Lambda air, aperture: 0.75). The Nikon NIS Elements software was used to operate the microscope and four distinct positions were acquired per well corresponding to overlays of the fluorescently labelled nuclei and the fluorescently labelled actin cytoskeleton (as a control cytoplasmic marker).

Images of fluorescently labelled nuclei were acquired as well as brightfield images from the same fields of view of four distinct cell lines (Table 1). The number of images per cell line was set to 240 for each. The fluorescent images were used to generate automatically the ground truth as described by Laufer *et al.* [12] using the EBImage R package [13]. Different cell lines were selected to cover as much as possible various cellular phenotypes (Figure 2) and morphologies which would allow us to generate a method that can be used on as many cell models as possible.

| Cell line | Origin | Cell type | # of images in dataset |
|---|---|---|---|
| A549 | Lung carcinoma | Epithelial | 240 |
| HeLa | Cervix adenocarcinoma | Epithelial | 240 |
| MCF7 | Breast adenocarcinoma | Epithelial | 240 |
| RPE1 | Retina (immortalized) | Epithelial | 240 |

**Table 1: Panel of selected cell lines**
Origin, name and cell type of the different cell lines tested in the study are given, information about the different cell lines are coming from the ATCC database [14].

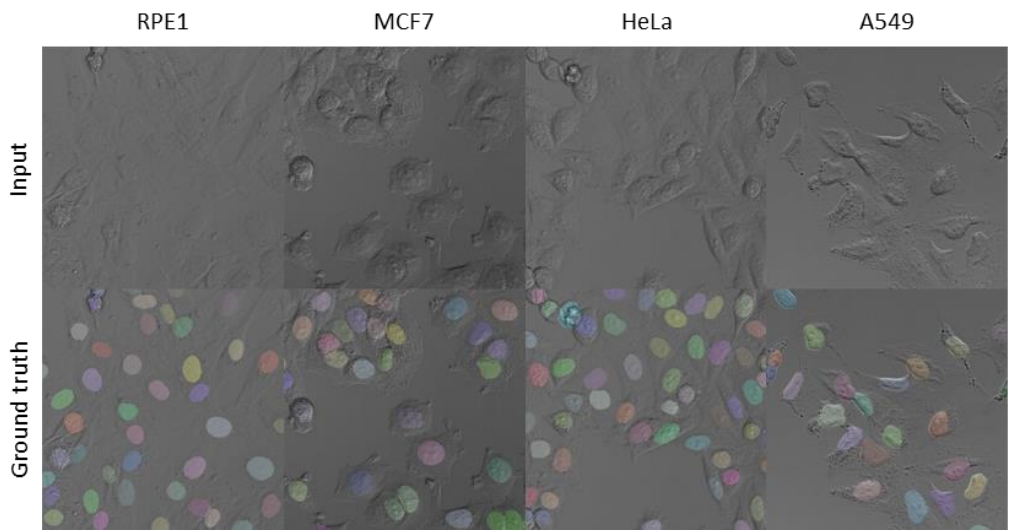

**Figure 2: Representative images of cell lines tested**
Input panel: Representative images of RPE1, MCF7, HeLa, A549 cells used in this study. Ground truth panel: overlay of ground truth instances obtained from fluorescently labeled nuclei displayed in various colors on top of the input image.

Often in the biomedical domain, the size of datasets is limited and therefore data augmentation strategy has proven to be efficient in computer vision [15]. To counterbalance the limited dataset size we had, we implemented a series of heavy data augmentation steps, using both geometry-based (rotations, flip, cropping and scaling) and intensity-based augmentations (contrast, brightness, gaussian blur, gaussian noise, gamma correction). We tested geometry-based operations alone or in combination with intensity-based operations (Table 2). The frequency and parameters of augmentation used are summarized in Figure 1Table 2.

| Operation | Parameters | Frequency |
| --- | --- | --- |
| Rotation (degrees) | 0; 90; 180; 270 | 1/1 |
| Flip | Horizontal - Vertical | For each value 1/2 |
| Cropping and scaling | 1.0 – 1.4 | 1/1 |
| No intensity-based operation | / | 1/6 |
| Contrast | 0.5 – 1.5 | 1/6 |
| Brightness | 0.5 – 1.5 | 1/6 |
| Gaussian blur | 1.0 – 1.5 | 1/6 |
| Gaussian noise | 0.1 | 1/6 |
| Gamma correction | 0.5 – 2.0 | 1/6 |

**Table 2: Data augmentation details**
All augmentation operations performed during training and their corresponding parameters and frequency of application are displayed here. For intensity-based operation only one was applied at the time whereas for geometry-based multiple operations were performed simultaneously.

### 2.2 Architectures

Two different CNN architectures were used in this study, U-Net [3] and Feature Pyramid Network (FPN) [7]. U-Net was chosen because it is the most commonly used architecture in the biomedical image segmentation field [4][8][10][16], its versatility and robustness makes it a very strong candidate for semantic segmentation. FPN was chosen because it achieved the best performance at the Data Science Bowl 2018 [5], challenge aiming at performing instance segmentation of nuclei from both labelled fluorescence and histological images. Both U-Net and FPN are encoder-decoder type of architectures. Within their encoders, stepwise convolutions are performed followed by non-linearity (ReLU) and pooling operations. The encoder/decoder pairing works as follow: the encoder, or the semantic path, decreases the resolution while leading to increased feature richness and global representation. This is then followed by the decoder, or geometric path, that uses skip connections [17] to combine the feature reach products of the encoder at the previous resolution stage with more convolutions performed after an upsampling step. Gradually, and with more convolutions and upsampling, the decoder generates higher resolution, finer and more precise probability maps to ultimately reach the original size of the input image in the final semantic map output. We described in detail the U-Net architecture in Figure 3 and FPN architecture in Figure 4.

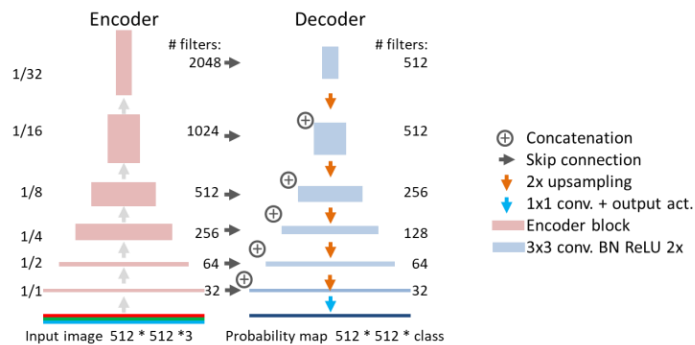

**Figure 3: U-Net architecture**
Architecture of U-Net used in this study. As described by Ronneberger *et al.* [3] it is composed of an encoder and a decoder. The network used in this study for plain U-Net was 7 levels deep whereas when using pre-trained encoders only 5 levels deep. The convolution block used was composed of convolution layer followed by batch normalization and ReLU activation (instance normalization and Leaky-ReLU in the case of nnU-Net implementation). The convolution block was repeated twice both in the encoder and decoder at each resolution stage.

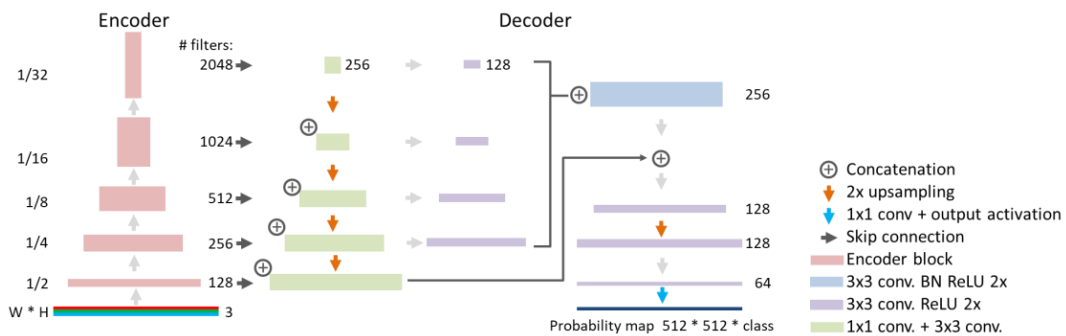

**Figure 4: Feature pyramid network architecture**
Architecture of FPN used in this study. As described by [7] it is composed of an encoder and a decoder. The base of this architecture is a U-Net with additional lateral branches using a combination of 1x1 and 3x3 convolutions in contrast to the original U-Net that uses only 3x3 convolutions.

## 2.3 Training, loss and final evaluation metric

### 2.3.1 Dice coefficient

The final evaluation metric used in this study is the Dice coefficient.

$$Dice_{Coefficient} = \frac{2\,|Y \cap Yhat|}{|Y| + |Yhat|} = \frac{2*TP}{2*TP + FP + FN}$$

Where Y corresponds to the ground truth and Yhat to the prediction. TP corresponds to the true positive, FP to the false positive and FN to the false negative.

### 2.3.2 Loss function

The combination of Dice coefficient and cross entropy was proven to be an efficient loss function for semantic segmentation tasks [4][18]. Using a loss function distinct from the Dice coefficient was shown to prevent over tuning towards the final evaluation metric [4]. For this reason, we used a loss function that combining both cross entropy and Dice coefficient.

$$Loss = cross\ entropy * 0.5\ +\ Dice_{Coeficient} * 0.5$$

### 2.3.3 Training

Training was done on a standard computer vision-oriented workstation (details listed in Table 9). To train the different CNN, parameters were kept constant except when changing deep learning frameworks (necessary for benchmarking against nnU-Net). All parameters used during the training are listed in Table 3. Parameter tuning was done in TensorFlow and using the A549 cell line dataset. The tuned parameters were then used on the other datasets and deep learning framework.

| Parameter / DL Framework | TensorFlow | PyTorch |
| --- | --- | --- |
| Epoch | 200 | 200 |
| Minibatch | 50 | 250 |
| Validation steps (per epochs) | Minibatch/5 = 10 | Minibatch/5 = 50 |
| Batch size | 8 | 11 |
| Patch size | 512 * 512 * 1 or 3 channels | 512 * 512 * 1 or 3 channels |
| Data augmentation | True (see table 3) | True (see table 3) |
| Learning rate | 1e-3 | 1e-3 (Adam) or 1e-2 (SGD) |
| Learning rate decay | Polynomial | Polynomial |
| Output layer activation | Softmax | Softmax |
| Optimizer | Adam | Adam or SGD |
| Dropout rate | 0.0 | 0.0 |

**Table 3: Network training.**
> For each deep learning frameworks investigated for training U-Net or FPN, we list the detailed hyper-parameters used. Adam optimizer was used as default optimizer. Only in the case of training plain U-Net in PyTorch, SGD with momentum was used as it gave rise to better results.

As the nnU-Net method uses SGD as optimizer for plain U-Net in PyTorch, we left this parameter untouched and simply changed the optimizer to Adam when using our U-Net with pre-trained encoder within the nnU-Net architecture. In TensorFlow the use of Adam increased the Dice score with every CNN tested including plain U-Net (data not shown). The batch size for the two deep learning frameworks was set to use optimally the GPU memory and the amount of minibatches was set to contain a full training time below eight hours using a standard GPU.

### 2.4 Instance segmentation from semantic maps

For the retrieval of individual nuclei two different instance segmentation strategies were tested. The first strategy was based on watershed [2]. As described in Figure 5 the output of the CNN will be one semantic map corresponding to entire nuclei. The semantic map is then thresholded, resulting objects are eroded, and watershed is used to separate adjacent objects and smaller objects are grown into the original mask again to form complete instances.

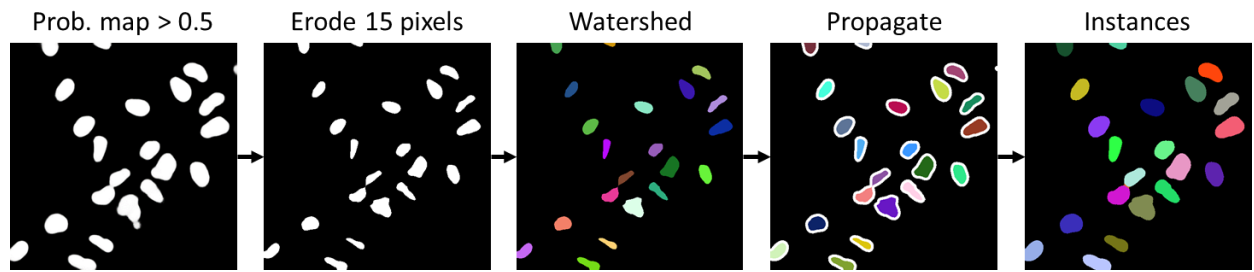

**Figure 5: Instance segmentation strategy using watershed.**
> Semantic maps predicted by the CNN are thresholded, then the resulting objects are eroded, watershed is applied on them before propagating the separated instances in the originally thresholded masks.

The second strategy uses as input two distinct semantic maps, one encoding for the centre and the other for the border of the nuclei. The centres are grown to their respective borders to generate complete nuclei [19].

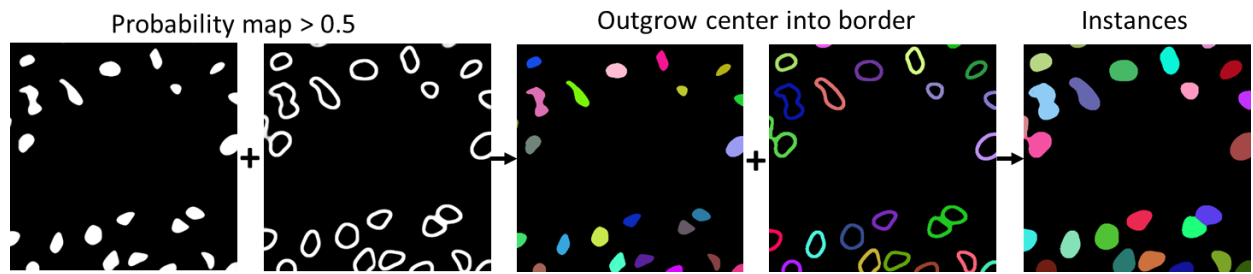

**Figure 6: Instance segmentation strategy using connected component analysis.**

Both semantic maps predicted by the CNN, centre and border, are thresholded than centre instances are gown into their neighbouring border regions forming complete instances.

## 3  RESULTS

To validate our architecture on a microscopy dataset, we tested a large sample of pre-trained encoders with the FPN architectures: ResNet, ResNeXt, Inception Resnet v2, Xception, DenseNet, NASNet mobile, EfficentNet B4 and Squeeze-and-Excitation network on the A549 dataset (Figure 7). The Squeeze-and-Excitation [20] family of network performed better with mean Dice score for SE ResNeXt 50 and 101 even above 0.91 while all other pretrained encoder produced values below 0.90.

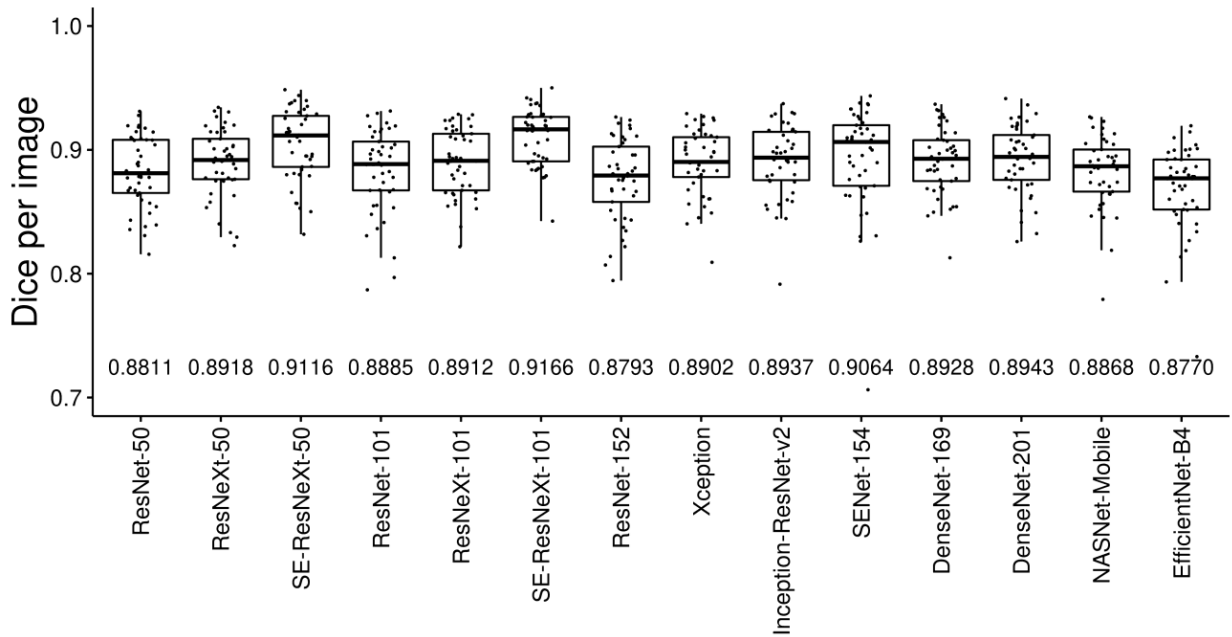

**Figure 7: Comparison of deep pre-trained encoders used in combination with the FPN architecture**
Mean Dice score per image from A549 test set obtained with FPN trained on the A549 dataset with various pretrained encoders. Average of mean Dice score per image for each encoder is displayed at the bottom of the panel.

As SE ResNeXt networks performed best with FPN, and in order to identify the best architecture, we compared SE ResNeXt 101 using either U-Net or FPN. We also compared SE ResNeXt 101 in both architectures to its predecessors, namely ResNeXt 101 and ResNet 101 (Figure 8). From this comparison it appears clearly that the channel wise attention mechanism from Squeeze-and-Excitation block increases the performance by at least 0.03 in Dice score for both architectures compared to ResNet 101 and 0.02 compared to ResNeXt 101. The results of SE ResNeXt 101 for both architectures were very close, with only 0.001 difference.

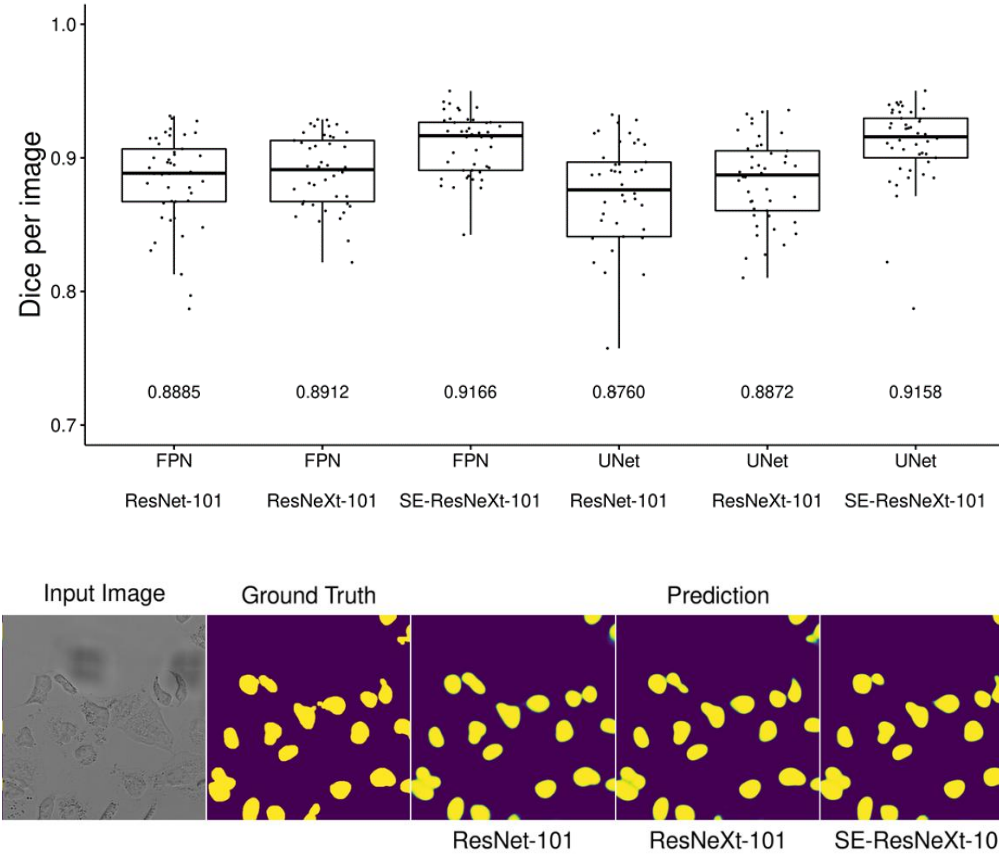

**Figure 8: Pre-trained encoder comparison of ResNet 101, ResNeXt 101 or SE ResNeXt 101 used in combination with U-Net**

Upper panel: Mean Dice score per image from A549 test set obtained with either FPN or U-Net trained on the A549 dataset with various pretrained encoders. Lower panel: one representative brightfield image with its corresponding nuclei ground truth and prediction from U-Net in combination with the indicated pre-trained encoder.

Because the results between U-Net and FPN using SE ResNeXt 101 as pre-trained encoder were very similar, we trained the models again using this time five-fold cross validation (Table 4) and eventually U-Net SE ResNeXt 101 performed best, 0.004 higher Dice score than FPN ensemble predictions (mention values). Not only the results from the ensemble prediction were higher than for its FPN counterpart, but also the inter-fold variability was lower, 0.003 and 0.008 standard deviation for U-Net and FPN respectively. For this reason, we concluded that U-Net SE ResNeXt 101 is the best performing CNN tested in this study for semantic segmentation of nuclei in brightfield images.

| Deep learning framework | | TensorFlow | TensorFlow |
| --- | --- | --- | --- |
| Architecture | | FPN | U-Net |
| Encoder | | SE ResNeXt 101 | SE ResNeXt 101 |
| Cell line | | A549 | A549 |
| Average Dice | (5-fold CV) | 0.912 | 0.915 |
| Standard-deviation | (5-fold CV) | 0.008 | 0.003 |
| Ensemble prediction | (5-fold CV) | **0.909** | **0.913** |

**Table 4: SE ResNeXt 101 as pre-trained encoder with U-Net or FPN.**
> Average of mean Dice score per image and corresponding standard deviation is given for the five folds of the five-fold cross validation procedure for either U-Net or FPN with SE ResNeXt 101. Additionally, average of mean Dice score per image is given for the ensemble prediction of all models coming from the five-fold cross validation procedure.

Since the training datasets in biology are often limited in terms of size, we investigated the influence of data augmentation strategies on the sematic segmentation performance. Data augmentation already proved to be an efficient strategy to increase performance and obtain more generic and broad-usage models [21]. For this reason, we tested two distinct strategies using either geometry-based augmentation alone or a combination of geometry-based and intensity-based augmentation (Table 5). We also examined dataset compositions, where we trained models using either microscopic images obtained from one cell line or from all the cell lines used in this study. In order to take into account, the Z-plan effect in this approach, we also trained models with images from either one Z-plane or from three Z-planes. The introduction of geometry-based augmentation increased the Dice score on average by 0.03 for the four cell lines tested whereas intensity-based augmentation increased only marginally the performance on Dice score by only 0.002. The introduction of more cell lines in the training set decreased the Dice score by 0.03, indicating that training with a generalist dataset was not beneficial for the overall performance of our model. On the other hand, introducing multiple Z-planes improved the segmentation results where we obtained a 0.01 increase in Dice score, probably due to the introduction of extra information from the other two additional Z-planes.

| | | | | | | | |
|---|---|---|---|---|---|---|---|
| Geometry-based augmentation | - | + | + | + | + | + | + |
| Intensity-based augmentation | - | - | - | + | - | - | + |
| Z-planes | 1 | 1 | 1 | 1 | 3 | 3 | 3 |
| # Cell lines used for training | 1 | 1 | 4 | 1 | 1 | 4 | 1 |
| A549 | 0.876 | 0.908 | 0.883 | 0.905 | 0.914 | 0.890 | 0.917 |
| HeLa | 0.888 | 0.907 | 0.870 | 0.910 | 0.915 | 0.894 | 0.912 |
| MCF7 | 0.898 | 0.908 | 0.865 | 0.910 | 0.914 | 0.876 | 0.917 |
| RPE1 | 0.835 | 0.897 | 0.899 | 0.903 | 0.915 | 0.931 | 0.919 |

**Table 5: Influence of data augmentation and dataset content on segmentation performance**
> Average of mean Dice score per image from the predictions of FPN SE ResNeXt 101 on the test set of the indicated cell lines is given.

After proving that our model could predict nuclei positions from brightfield images in a robust and reliable way, we decided to compare our method using U-Net SE ResNeXt 101 developed with TensorFlow framework to two different studies. The first one from Fishman *et al.* [6] where they focused on semantic segmentation of nuclei in brightfield images and second to nnU-Net [4], one of the best performing and most commonly used method in the biomedical domain. As Fishman *et al.* did not provide with data or source code we attempted to replicate the results of their work (Table 6). In order to fairly compare to nnU-Net we ported our model, U-Net SE ResNeXt 101, to PyTorch and implemented it to the nnU-Net method. This operation was done to prevent false conclusion coming from inter deep learning framework comparison.

When attempting to replicate the results of Fishman *et al.* with plain U-Net and our data it appeared that our implementation out-performed theirs on Dice score on average by 0.06 (Table 6). It is important to remind that we could not replicate completely their results since neither the source code, nor the images are available. Nonetheless our implementation performed better with the images we generated. When replacing plain U-Net with U-Net SE ResNeXt 101 the performance increased on average by 0.02 in Dice score. We then compared our method with plain U-Net from nnU-Net which displayed on average a Dice score 0.02 higher. As we mentioned earlier the last comparison is not totally fair as deep learning frameworks varies, for this reason when implementing our model into nnU-Net, we could improve the results from standard nnU-Net by 0.01 Dice score on average for the four cell lines tested, proving that the pre-trained encoder SE ResNeXt 101 improves the semantic segmentation of nuclei in brightfield images with another semantic segmentation implementation than ours.

| Method | Fishman | Ours | Ours | nnU-Net | Ours + nnU-Net |
| --- | --- | --- | --- | --- | --- |
| DL framework | Unknown | TensorFlow | TensorFlow | PyTorch | PyTorch |
| Architecture | U-Net | U-Net | U-Net | U-Net | U-Net |
| Pre-trained encoder | / | / | SE101 | / | SE101 |
| Data | Fishman | Ours | Ours | Ours | Ours |
| A549 | 0.815 | 0.885 | 0.913 | 0.926 | 0.937 |
| HeLa | 0.851 | 0.889 | 0.905 | 0.927 | 0.937 |
| MCF7 | 0.805 | 0.891 | 0.903 | 0.923 | 0.936 |
| RPE1 | / | 0.883 | 0.906 | 0.926 | 0.936 |

**Table 6: Benchmarking**
For each cell line and architecture tested, we provide with the average of mean Dice score per image. Predictions were obtained from an ensemble of U-Net SE ResNeXt 101 models trained using five-fold cross validation. Results for the Fishman method were ported from [6].

In order to avoid an over tuning bias towards Dice score, we also evaluated our CNN model using other metrics. Results in Table 8 show that not only Dice score was increased with our model in comparison to plain U-Net using the nnU-Net method, but also all other metrics improved (accuracy, precision, recall).

In order to determine which of plain U-Net or U-Net SE ResNeXt 101 produced the best instance segmentation results, we tested our two instance segmentation strategies on the semantic maps generated by the two CNN models. Focusing on mean Dice score per instance, both for plain U-Net and U-Net SE ResNeXt 101, the connected component analysis performed better than the watershed-based strategy. In addition, between U-Net with SE ResNeXt 101 and plain U-Net, the first one performed better for both instance segmentation method tested. U-Net SE ResNeXt 101 with connected component analysis was the combination that generated the highest mean Dice score at the instance level.

| Method | Architecture | Encoder | Post-processing | Mean Dice score per instance |
| --- | --- | --- | --- | --- |
| nnU-Net | U-Net | / | Watershed | 0.874 |
| nnU-Net | U-Net | / | CCA | 0.883 |
| nnU-Net | U-Net | SE ResNeXt 101 | Watershed | 0.883 |
| nnU-Net | U-Net | SE ResNeXt 101 | CCA | 0.893 |

**Table 7: Dice score for each instance segmentation strategy**
Mean Dice score per instance is given for each listed combination. The semantic maps were generated by combining the CNN architecture with the encoder. Predictions from ensemble models trained using five-fold cross validation, were then post processed with the indicated post-processing.

To ensure proper instance segmentation quality, we used the approach described by Caicedo et al. [8]. We investigated the number of instances generated in comparison to the ground truth, but also looked at extra introduced objects, as well as missed objects in addition to over- and under-split objects (Figure 9).

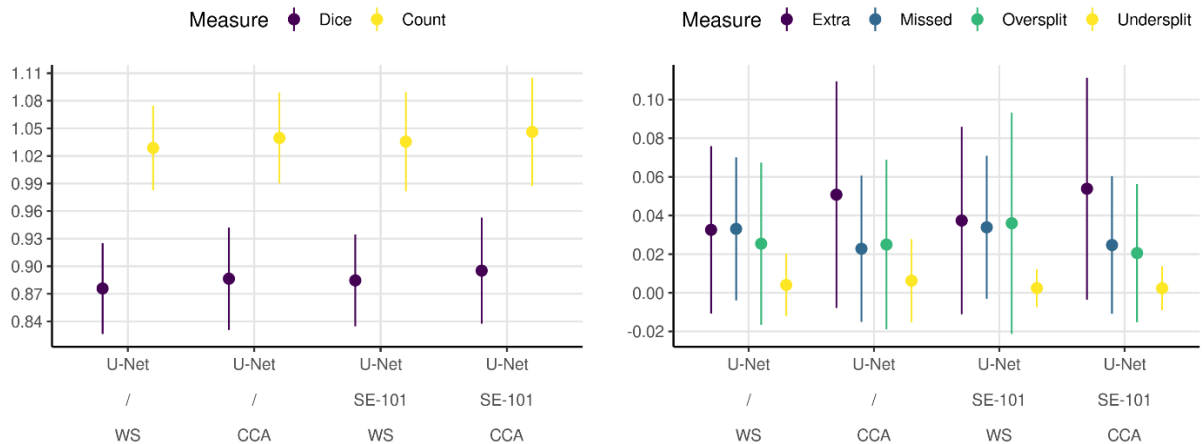

**Figure 9: Detailed instances evaluation metrics**
>Left panel: Mean and standard deviation of mean Dice coefficient per instance per image and mean predicted cell count relative to ground truth cell count per image. Right panel: Mean and standard deviation of extra, missed, under- or over-split objects in comparison to ground truth are displayed.

None of the combinations of network models and instance segmentation strategy generated more than five percent extra objects or led to a decrease in the number of instances. From the representative images we could conclude that the increase in extra or over-split objects comes from either better separated adjacent instances in comparison to ground truth or small instances detected at the rim of the image (Figure 11). Moreover, the use of connected component analysis enabled to better resolve clumped nuclei regions hence increasing further the amount of extra or over-split nuclei in comparison to ground truth, despite the instance being properly segmented. Under-split nuclei were almost absent in all conditions tested.

To complement our investigation on the quality of instances generated, we also generated samples closer to an actual biological application. We treated cells with one of those compounds: Staurosporin, (apoptosis inducer), Doxorubicin (DNA damage inducer), Nocodazole (microtubule destabilizing agent), Vorinostat (epigenetic modulator of histone acetylation), Nigericin (Ionophore). The diversity of these molecules was planned so that we targeted diverse biological mechanisms. In addition, we titrated the concentrations of these drugs to obtain a large dynamic range of cellular response and ensure a mild to a strong effect as reflected by the cell counts (Figure 10) and morphologically observed in Figure 13. In order to directly examine Dice coefficient at the image level and instance level, we used watershed-based instance segmentation strategy. Additionally, we trained our CNN model using images from vehicle-treated only samples.

In almost all the conditions tested, there was no difference observed between the predicted and the ground truth cell counts (Figure 10), although we used only vehicle treated samples as a training set, proving the method can accurately predict nuclei in diverse biological context. Moreover, and despite the drastic drop in cell count, the number of nuclei predicted was almost identical to the ground truth. The only exception was in the two highest doses of doxorubicin where a discrepancy was observed with the ground truth. When examining this difference between the ground truth and the predicted values we noticed that the reason was that the intensity of fluorescently labelled nuclei in very high doses of doxorubicin decreased severely (Figure 14) preventing proper ground truth generation. In fact, our model did segment the instances from the brightfield images as expected (Figure 13), proving the robustness of the method proposed here but the ground truth, used as a benchmark, was not accurate enough.

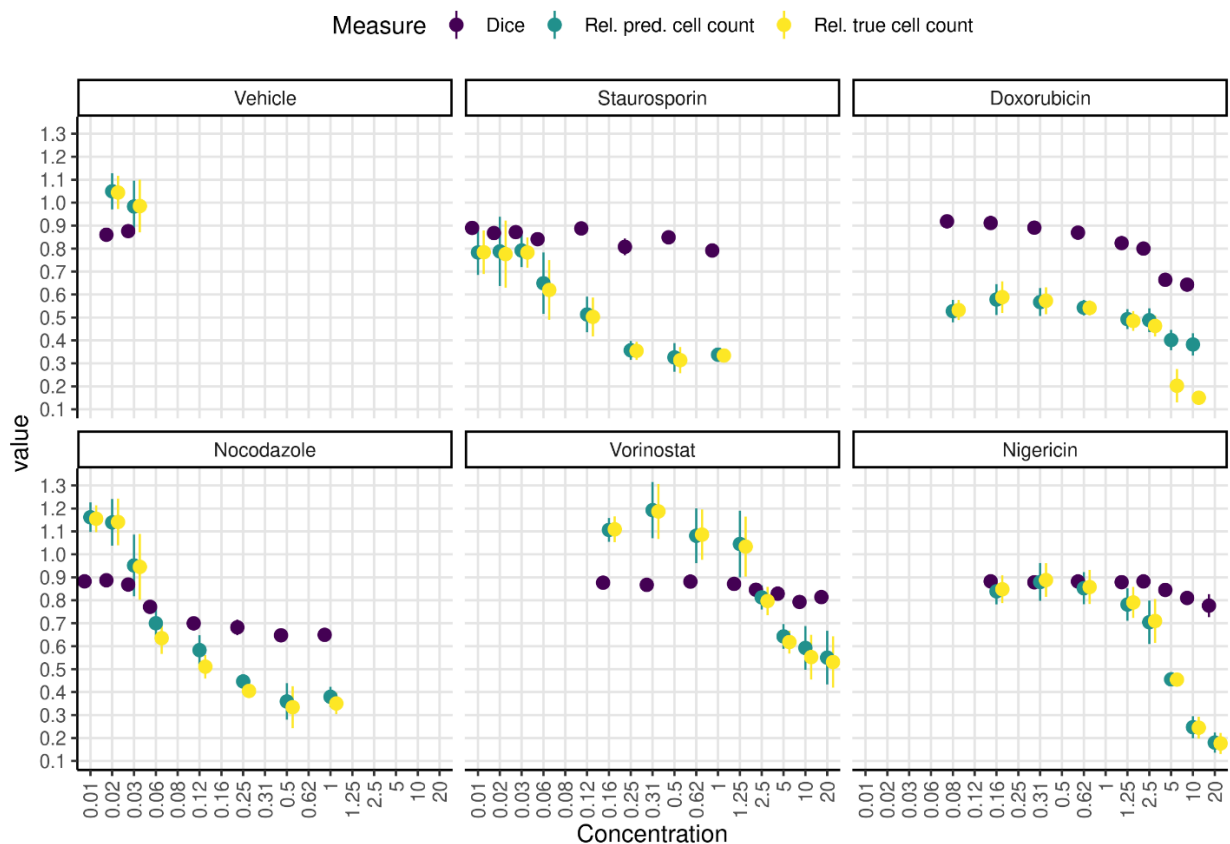

**Figure 10: Detailed Dice score and cell count for instances generated by watershed strategy upon treatment with small molecule inhibitors**

> For each small molecule inhibitor tested, mean and standard deviation of mean Dice coefficient per instance per image and mean predicted or ground truth cell count relative to vehicle control treated samples per image are displayed. All results shown came from U-Net SE ResNeXt 101 trained on vehicle treated samples only using five-fold cross validation procedure. Watershed-based strategy was used to generate the instances from the semantic maps coming from the ensemble predictions.

Dice coefficient was observed to decrease with decreasing viability, and for this reason we hypothesized that the CNN model being trained on images from vehicle treatment only, could not generalize as well on images from samples diverging from the negative control. For instance, Nocodazole treatment led to the strongest decrease in Dice score among all compounds tested, whereas on the other hand Nigericin treatment, although decreasing cell count stronger at high doses than Nocodazole, did not affect Dice score. Looking closer at Nocodazole (Figure 14) it appeared that out of focus nuclei could be found already with treatment at 125 nM and above. The loss of focus is due to changes in morphology that was not observed as dramatically with other compounds and explain the decrease in Dice score. We also looked at extra, missed, over- and under-split objects (Figure 12), the observations followed the comments given for cell count and Dice score.

## 4 DISCUSSION

In this study we identified U-Net with SE ResNeXt 101 as the best combination of CNN architecture and pre-trained encoder for semantic segmentation of nuclei in brightfield images, with Dice score of 0.94 on average for four distinct cell lines when combined with the nnU-Net method. We also provided two distinct instance segmentation methods using as input the semantic maps generated by a CNN model that we initially optimized. The best instance segmentation method gave us a Dice score per object of up to 0.89 for the A549 cell line dataset.

From our investigation, deep pre-trained encoders with Squeeze-and-Excitation blocks[20] improved the performance of U-Net and FPN the most, therefore deeper investigation with Squeeze-and-Excitation blocks could be of great interest especially when incorporated in plain U-Net architecture. It could alleviate a lightweight architecture faster to train and run inference with. In addition, we showed that pre-trained encoders incorporating SE blocks and residual connections produced the highest Dice scores [23], and therefore, it could be interesting to train a modified plain U-Net architecture with residual connections and Squeeze-and-excitation block, on the level of the encoder, the decoder or both.

Additionally, we suggest that more investigation in the structure of the training set could be beneficial to improve the segmentation of nuclei in brightfield images. Despite our results indicating that a specialized model performs better than a model trained on a mixed data set, other studies [24] suggest that having a more diverse data set and eventually training longer the network could help generalizing better and hence improving the segmentation. Moreover, incorporating images from various biological contexts and treatments, as done here with the small molecule inhibitors treatments, would also enable to encompass more phenotypic diversity and hence help to further improve segmentation.

We also validated our method on brightfield images from cell lines treated with small molecule inhibitors, proving that our method enables the analysis of actual biological samples in a more biologically relevant context. For instance, cytotoxicity assays would be greatly facilitated with our method as we showed that predicted and ground truth cell counts are almost identical. Eventually, and for applications combining fluorescence microscopy, our method enables to free the usual nuclear stained channel so that it can be used for more biologically relevant markers.

# 5 REFERENCES

[1] C. Laufer, B. Fischer, M. Billmann, W. Huber, and M. Boutros, "Mapping genetic interactions in human cancer cells with RNAi and multiparametric phenotyping.," *Nat. Methods*, vol. 10, no. 5, pp. 427–31, 2013, doi: 10.1038/nmeth.2436.

[2] N. Malpica *et al.*, "Applying watershed algorithms to the segmentation of clustered nuclei," *Cytometry*, vol. 28, no. 4, pp. 289–297, 1997, doi: 10.1002/(SICI)1097-0320(19970801)28:4<289::AID-CYTO3>3.0.CO;2-7.

[3] O. Ronneberger, P. Fischer, and T. Brox, "U-Net: Convolutional Networks for Biomedical Image Segmentation," 2015, pp. 234–241.

[4] F. Isensee, P. F. Jaeger, S. A. A. Kohl, J. Petersen, and K. H. Maier-Hein, "nnU-Net: a self-configuring method for deep learning-based biomedical image segmentation," *Nat. Methods*, vol. 18, no. 2, pp. 203–211, Feb. 2021, doi: 10.1038/s41592-020-01008-z.

[5] J. C. Caicedo *et al.*, "Nucleus segmentation across imaging experiments: the 2018 Data Science Bowl," *Nat. Methods*, Oct. 2019, doi: 10.1038/s41592-019-0612-7.

[6] D. Fishman *et al.*, "Practical segmentation of nuclei in brightfield cell images with neural networks trained on fluorescently labelled samples," *J. Microsc.*, vol. 284, no. 1, pp. 12–24, Oct. 2021, doi: 10.1111/jmi.13038.

[7] S. Seferbekov, V. Iglovikov, A. Buslaev, and A. Shvets, "Feature pyramid network for multi-class land segmentation," *IEEE Comput. Soc. Conf. Comput. Vis. Pattern Recognit. Work.*, vol. 2018-June, pp. 272–275, 2018, doi: 10.1109/CVPRW.2018.00051.

[8] J. Caicedo *et al.*, "Evaluation of Deep Learning Strategies for Nucleus Segmentation in Fluorescence Images," *bioRxiv*, p. 335216, 2018, doi: 10.1101/335216.

[9] S. M. A. Hayajneh, M. H. Alomari, and B. Al-shargabi, "Cascades Neural Network based Segmentation of Fluorescence Microscopy Cell Nuclei," vol. 9, no. 5, pp. 275–285, 2018.

[10] V. Iglovikov and A. Shvets, "TernausNet: U-Net with VGG11 Encoder Pre-Trained on ImageNet for Image Segmentation," Jan. 2018.

[11] R. Ribani and M. Marengoni, "A Survey of Transfer Learning for Convolutional Neural Networks," *Proc. - 32nd Conf. Graph. Patterns Images Tutorials, SIBGRAPI-T 2019*, pp. 47–57, 2019, doi: 10.1109/SIBGRAPI-T.2019.00010.

[12] C. Laufer, B. Fischer, M. Billmann, W. Huber, and M. Boutros, "Mapping genetic interactions in human cancer cells with RNAi and multiparametric phenotyping," *Nat. Methods*, vol. 10, no. 5,

pp. 427–431, 2013, doi: 10.1038/nmeth.2436.

[13] G. Pau, F. Fuchs, O. Sklyar, M. Boutros, and W. Huber, "EBImage-an R package for image processing with applications to cellular phenotypes," *Bioinformatics*, vol. 26, no. 7, pp. 979–981, 2010, doi: 10.1093/bioinformatics/btq046.

[14] "American Type Culture Collection." [Online]. Available: https://www.atcc.org/.

[15] C. Shorten and T. M. Khoshgoftaar, "A survey on Image Data Augmentation for Deep Learning," *J. Big Data*, vol. 6, no. 1, 2019, doi: 10.1186/s40537-019-0197-0.

[16] T. Falk *et al.*, "U-Net: deep learning for cell counting, detection, and morphometry," *Nat. Methods*, vol. 16, no. 1, pp. 67–70, 2019, doi: 10.1038/s41592-018-0261-2.

[17] M. Drozdzal, E. Vorontsov, G. Chartrand, S. Kadoury, and C. Pal, "The importance of skip connections in biomedical image segmentation," *Lect. Notes Comput. Sci. (including Subser. Lect. Notes Artif. Intell. Lect. Notes Bioinformatics)*, vol. 10008 LNCS, pp. 179–187, 2016, doi: 10.1007/978-3-319-46976-8_19.

[18] S. Seferbekov, A. Buslaev, and D. Victor, "DSB2018 [ods.ai] topcoders 1st place solution," *GitHub*, 2018. [Online]. Available: https://github.com/selimsef/dsb2018_topcoders.

[19] F. Isensee, P. F. Jaeger, S. A. A. Kohl, J. Petersen, and K. H. Maier-Hein, "Cell Tracking Challenge - DKFZ-GE Authors Submission," vol. 1, no. 1, pp. 1–3, 2020.

[20] J. Hu, L. Shen, S. Albanie, G. Sun, and E. Wu, "Squeeze-and-Excitation Networks," *IEEE Trans. Pattern Anal. Mach. Intell.*, vol. 42, no. 8, pp. 2011–2023, 2020, doi: 10.1109/TPAMI.2019.2913372.

[21] N. E. Khalifa, M. Loey, and S. Mirjalili, "A comprehensive survey of recent trends in deep learning for digital images augmentation," *Artif. Intell. Rev.*, vol. 55, no. 3, pp. 2351–2377, 2022, doi: 10.1007/s10462-021-10066-4.

[22] C. A. Belmokhtar, J. Hillion, and E. Ségal-Bendirdjian, "Staurosporine induces apoptosis through both caspase-dependent and caspase-independent mechanisms," *Oncogene*, vol. 20, no. 26, pp. 3354–3362, 2001, doi: 10.1038/sj.onc.1204436.

[23] K. He, X. Zhang, S. Ren, and J. Sun, "Deep Residual Learning for Image Recognition," in *2016 IEEE Conference on Computer Vision and Pattern Recognition (CVPR)*, 2016, vol. 2016-Decem, pp. 770–778, doi: 10.1109/CVPR.2016.90.

[24] C. Stringer, T. Wang, M. Michaelos, and M. Pachitariu, "Cellpose: a generalist algorithm for cellular segmentation," *Nat. Methods*, vol. 18, no. 1, pp. 100–106, Jan. 2021, doi: 10.1038/s41592-020-01018-x.

# 6 SUPPLEMENTARY MATERIAL

| Method | nnU-Net | | | | | | | |
|---|---|---|---|---|---|---|---|---|
| Architecture | U-Net | | | | | | | |
| Pre-trained encoder | / | SE101 | / | SE101 | / | SE101 | / | SE101 |
| Cell line | A549 | A549 | HeLa | HeLa | MCF7 | MCF7 | RPE1 | RPE1 |
| Accuracy | 0.982 | 0.985 | 0.973 | 0.976 | 0.982 | 0.984 | 0.975 | 0.978 |
| Dice coefficient | 0.926 | 0.936 | 0.927 | 0.937 | 0.923 | 0.932 | 0.926 | 0.936 |
| False Discovery Rate | 0.073 | 0.060 | 0.073 | 0.062 | 0.071 | 0.063 | 0.073 | 0.063 |
| False Negative Rate | 0.074 | 0.068 | 0.072 | 0.064 | 0.081 | 0.071 | 0.075 | 0.065 |
| False Omission Rate | 0.010 | 0.009 | 0.017 | 0.015 | 0.011 | 0.010 | 0.015 | 0.013 |
| False Positive Rate | 0.011 | 0.009 | 0.017 | 0.015 | 0.010 | 0.009 | 0.015 | 0.013 |
| Jaccard index | 0.864 | 0.880 | 0.865 | 0.881 | 0.858 | 0.873 | 0.863 | 0.880 |
| Neg. Predictive Value | 0.990 | 0.991 | 0.983 | 0.985 | 0.989 | 0.990 | 0.985 | 0.987 |
| Precision | 0.927 | 0.940 | 0.927 | 0.938 | 0.929 | 0.937 | 0.927 | 0.937 |
| Recall | 0.926 | 0.932 | 0.928 | 0.936 | 0.919 | 0.929 | 0.925 | 0.935 |
| Total Pos. Reference | 31364 | 31364 | 48857 | 48857 | 32007 | 32007 | 43901 | 43901 |
| Total Positives Test | 31509 | 31312 | 48917 | 48841 | 31838 | 31879 | 43798 | 43785 |
| True Negative Rate | 0.989 | 0.991 | 0.983 | 0.985 | 0.990 | 0.991 | 0.985 | 0.987 |

**Table 8: Detailed evaluation metrics from prediction of U-Net with or without SE ResNeXt 101**
Detail of evaluation metric for predictions from different cell lines using either plain U-Net or U-Net SE ResNeXt 101.

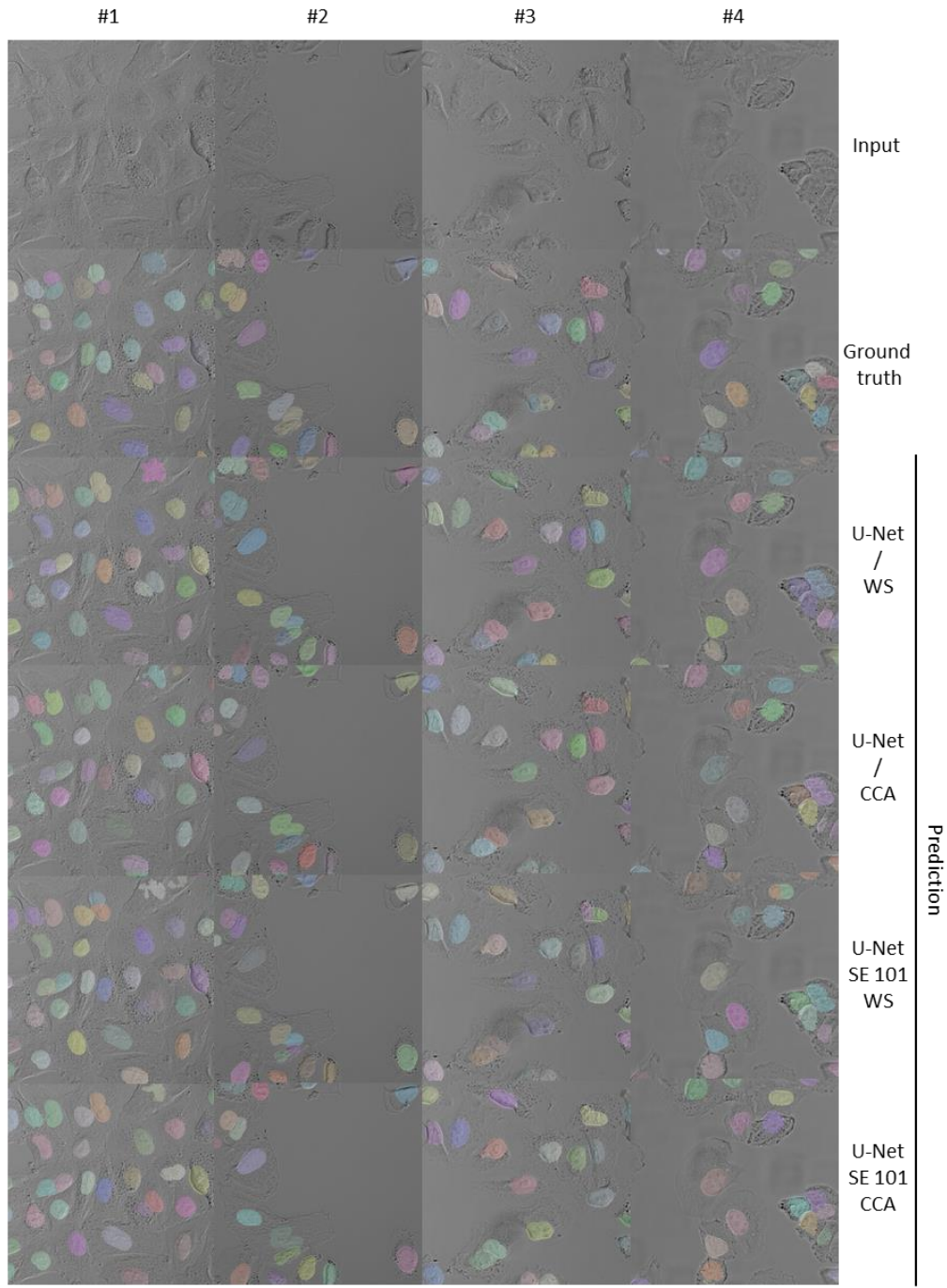

Figure 11: Representative instances generated with or without pre-trained encoder using one or two classes probability maps as input.

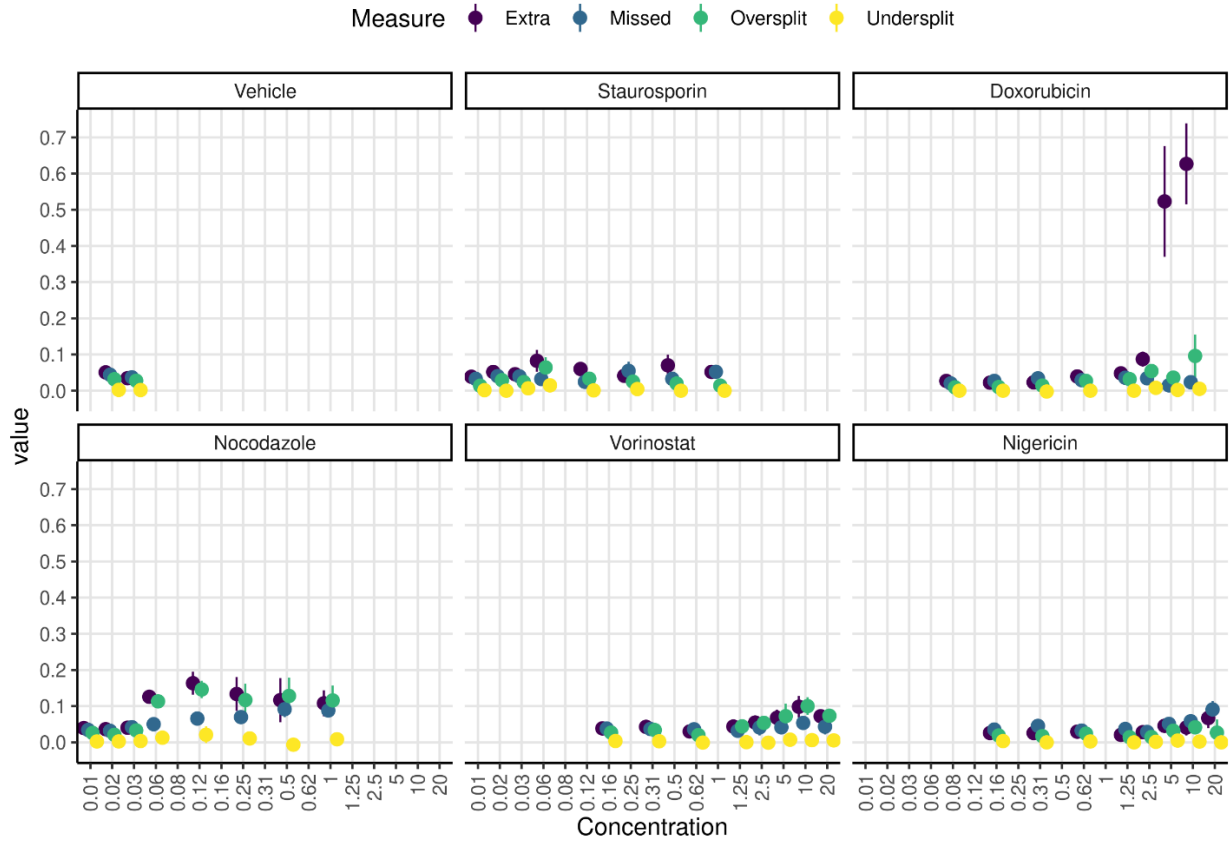

Figure 12: Detailed extra, missed, over-split and under-split objects counts for instances generated by watershed strategy upon treatment with different small molecule inhibitors at different concentrations

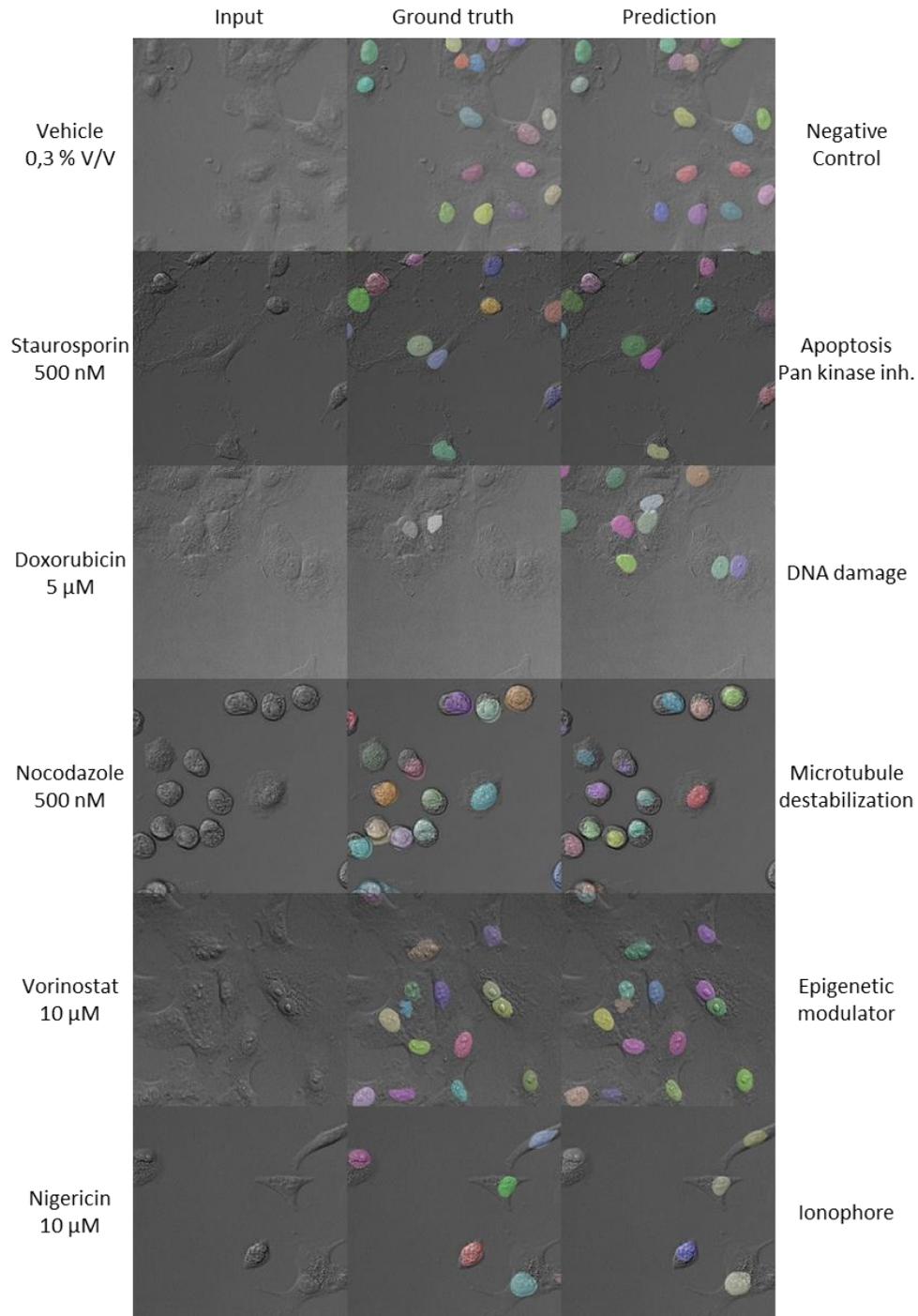

**Figure 13: Example instances generated by watershed strategy upon treatment with small molecule inhibitors**

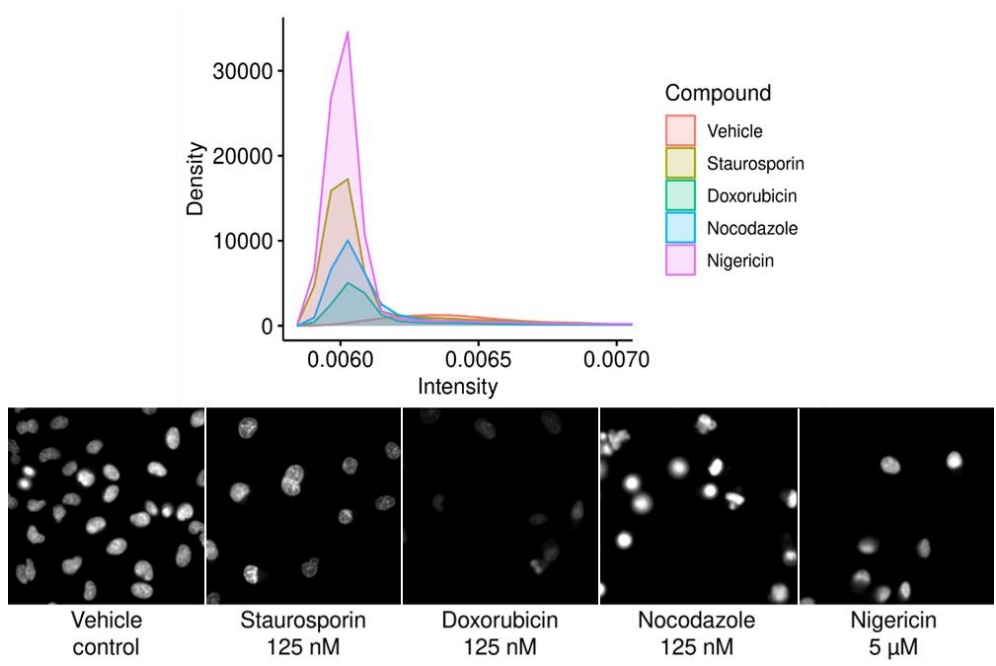

**Figure 14: Extreme phenotypes obtained upon treatment with small molecule inhibitors**

| | | | |
|---|---|---|---|
| HARDWARE | CPU | | 1X INTEL XEON CPU E5-1630 V3 |
| | GPU | | 1x Nvidia RTX 3090 |
| | RAM | | 256 GB |
| | Hard drive | | 1x Solid state drive 2 TB + 1x SATA 2TB |
| SOFTWARE & LIBRARIES | Operating system: Ubuntu | | 20.04 |
| | CUDA | | 11.2 |
| | cuDNN | | 8.1 |
| | Python | PyTorch | 1.9.0 |
| | Python | TensorFlow | 2.4.1 |
| | Python | Keras | 2.4.3 |
| | Python | nnU-Net | 1.7.0 |
| | R | TensorFlow | 2.4.0 |
| | R | Keras | 2.4.0 |
| | R | EBImage | 4.33.0 |

**Table 9: Specifics of hardware and software**